\documentstyle[aps]{revtex}
\begin{document}
\author{D. Foerster}
\address{CPTMB, URA1537\\
Universit\'{e} de Bordeaux I\\
Rue du Solarium,\\
33174 Gradignan, France}
\title{Can Majorana Fermions correctly describe the ordered state of an
antiferromagnetic Heisenberg chain? }
\maketitle

\begin{abstract}
To check the reliability of the Majorana representation of quantum spin
systems, we use it to compute the ground state energy of an
antiferromagnetic spin 1/2 chain to one loop order. We find a very small one
loop correction of the mean field energy and a discrepancy of $>100\%$
compared to the Bethe Ansatz result. We conclude that a careful handling of
the gauge degrees of freedom of this representation is crucial to get
correct results.

\end{abstract}

\section{Introduction}

The Majorana representation is a promising tool for describing frustrated
quantum spin 1/2 antiferromagnets and heavy fermion systems\cite{CMT},\cite
{CIT}. It provides a description of spin operators in terms of bilinears of
real fermions without any constraints. It is useful to consider the
representation of spin operators of the group $O(N)$ and take $N=3$ only at
the end: 
\begin{eqnarray}
O(N) &:&s_{x}^{kl}=-\frac{i}{2}\left( \eta _{x}^{k}\eta _{x}^{l}-\eta
_{x}^{l}\eta _{x}^{k}\right) \text{, }k,l=1,2,..N\text{, }\left[ \eta
_{x}^{k},\eta _{y}^{l}\right] _{+}=\delta _{kl}\delta _{xy}\text{, }\eta
_{x}^{k+}=\eta _{x}^{k}  \nonumber \\
O(3)\text{{}} &:&\text{ }\overrightarrow{s_{x}}=-\frac{i}{2}\overrightarrow{%
\eta _{x}}\times \overrightarrow{\eta _{x}}
\end{eqnarray}
A mean field treatment of spin systems in this representation gives rise to $%
O(N)$ invariant bond variables in a natural way and, therefore, this
representation seems suited for a description of the disordered or
''liquid'' phases of quantum spin systems. Indeed it is not difficult to
see, by using the Majorana representation, that most $O(N)$ spin systems (in
the lowest representation that generalizes spin $\frac{1}{2}$) reduce to RVB
like spin liquids at $N=\infty $ \cite{DFFT}. However, for the Majorana
representation to be really useful, it must also be able to describe the
competition between ordered and liquid spin systems as a function of the
amount of frustration and for $N=3$.

Recently, it was pointed out that a simple Hartree Fock treatment of the
spin 1/2 Heisenberg antiferromagnetic chain gives qualitatively correct
results\cite{Shastry}. The purpose of the present note is to show that in a
more systematic approach the one loop correction to the mean field energy is
small and that the ground state energy still differs from the exact Bethe
Ansatz result by more than 100\%.

It is very likely that the local $Z_{2}$ invariance of the Majorana
representation is responsible for this discrepancy and that it must be dealt
with correctly in order to obtain reliable results. The situation is
reminiscent of the Yang Mills theory of strong interactions in which a
correct treatment of the local invariance is crucial.

Although the local $Z_{2}$ invariance of the Majorana representation was
previously dealt with in the context of the two channel Kondo problem \cite
{CIT} a more systematic treatment of this invariance using standard gauge
theory techniques is needed.

\section{Mean field}

The Hamiltonian of the $O(N)$ antiferromagnet is

\begin{eqnarray}
O(N) &:&H=\sum_{<x,y>k,l}\frac{1}{2}s_{x}^{kl}s_{y}^{kl}-\frac{N}{8}=\frac{1%
}{2}\sum_{<x,y>}(\eta _{x\mu }\eta _{y\mu })^{2}  \nonumber \\
O(3) &:&H=\sum_{<x,y>}\overrightarrow{s}_{x}\cdot \overrightarrow{s}_{y}-%
\frac{3}{8}=\frac{1}{2}\sum_{<x,y>}(\overrightarrow{\eta }_{x}\cdot 
\overrightarrow{\eta }_{y})^{2}
\end{eqnarray}
with $<x,y>$ denoting neighboring points, $J=1$ and we will suppress the
internal symmetry indices in the following. To allow for a systematic
treatment of fluctuations the mean field must be introduced by a
Hubbard-Stratonovich transformation: 
\begin{eqnarray}
Z &=&Tre^{-\beta H}=\int D\eta e^{-\int_{0}^{\beta }dt\left( \frac{1}{2}\eta
_{\mu }\partial _{t}\eta _{\mu }+H(\eta )\right) }  \nonumber \\
&=&\int DBD\eta e^{-\int_{0}^{\beta }dt\left( \frac{1}{2}\eta _{\mu
}\partial _{t}\eta _{\mu }+H(B,\eta )\right) }\text{, }H(B,\eta )=\frac{1}{4}%
\sum_{x,y}\frac{B_{xy}^{2}}{J_{xy}}+\frac{i}{2}\sum_{x,y}B_{xy}\eta _{x}\eta
_{y}  \nonumber \\
&\rightarrow &Z=\int DBe^{-\beta F(B)}\text{, }F(B)=-\frac{1}{2\beta }\log
\det \left( \partial _{\tau }+iB\right) 
\end{eqnarray}
For a $d=1$ chain and with a uniform auxiliary field 
\begin{equation}
B_{xy}=\frac{M}{2}\left( \delta _{x,y+1}-\delta _{y,,x+1}\right) 
\end{equation}
we find 
\begin{eqnarray}
H(B,\eta ) &=&\frac{L}{8}M^{2}-\frac{M}{2}\sum_{p>0}\left( \eta _{p\mu }\eta
_{-p\mu }-\eta _{-p\mu }\eta _{p\mu }\right) \sin p  \nonumber \\
\left[ \eta _{p},\eta _{q}\right] _{+} &=&\delta _{p+q,0}
\end{eqnarray}
where $L$ denotes the length of the chain. The mean field saddle point is $M=%
\frac{2N}{\pi }$ and gives a quasiparticle dispersion and ground state
energy (for the original Heisenberg Hamiltonian) of 
\begin{eqnarray}
E(p) &=&\frac{2N}{\pi }\sin p\stackrel{N=3}{=}\frac{6}{\pi }\sin p\text{
(Bethe Ansatz: }\frac{\pi }{2}\sin p)  \nonumber \\
\frac{<H(B,\eta )>}{L}+\frac{N}{8} &=&-\frac{N^{2}}{2\pi ^{2}}+\frac{N}{8}%
\stackrel{N=3}{=}-0.08\text{ (Bethe Ansatz:}-0.44\text{)}
\end{eqnarray}
It is known that the optimal mean field in the Majorana representation is
completely dimerized for a large class of Heisenberg antiferromagnets in any
dimension \cite{DFFT} and we could check whether the fluctuations about the
mean field restore the uniform mean field configuration in $d=1$. However,
for our purpose of demonstrating a bug in an uncritical use of the Majorana
representation this is unnecessary since:

\begin{itemize}
\item  the exact Bethe Ansatz solution does not break translations -

\item  as we will see the fluctuations about the uniform mean field in the
Majorana representation give only a small correction to the mean field
energy, with the one loop corrected ground state energy still differing from
the exact result by more than 100\%.
\end{itemize}

\section{One loop correction}

The self consistent Gaussian approximation is a very reasonable
approximation that gives the first 3 terms of the Stirling series of $\Gamma
(n)=\int_{0}^{\infty }dxx^{n-1}e^{-x}$ at one stroke and which {\em should},
a priori, give good results also for the spin chain in the Majorana
representation. In this approximation, the partition function reduces to 
\begin{equation}
Z\sim \frac{e^{-\beta F(B)}}{\sqrt{\det \frac{\delta ^{2}F}{\delta B^{2}}}}%
\equiv e^{-\beta \widetilde{F}(B)}\text{, }\widetilde{F}(B)=F(B)+\frac{1}{%
2\beta }\log \det \frac{\delta ^{2}F}{\delta B^{2}}
\end{equation}
To identify the quadratic form $\frac{\delta ^{2}F}{\delta B^{2}}$one must
expand the free energy $F(B)$ about the saddle point 
\begin{eqnarray}
F(B+\delta B) &=&F(B)+\frac{1}{2}\frac{\delta ^{2}F}{\delta B^{2}}\delta
B\delta B+O\left( \delta B,\delta B^{3}\right)  \nonumber \\
\frac{1}{2}\frac{\delta ^{2}F}{\delta B^{2}}\delta B\delta B &=&\frac{1}{%
4\beta }\int_{0}^{\beta }d\tau \sum_{x,y}\delta B_{xy}^{2}+\frac{1}{4\beta }%
tr\left( \frac{1}{\partial _{t}+iB}\right) \delta B\left( \frac{1}{\partial
_{t}+iB}\right) \delta B
\end{eqnarray}
After some simplifications the total quadratic form turns out to be 
\begin{eqnarray}
\frac{\delta ^{2}F}{\delta B^{2}}\delta B\delta B &\sim &\sum \delta
b^{*}(q)F(q)\delta b(q)  \nonumber \\
F(q) &=&1+\frac{1}{L\beta }\sum_{p+r-q=0}\left( 1-e^{i(r-p)}\right) g(p)g(r)%
\text{, }q=(q_{0},q_{1})  \nonumber \\
\delta B &=&\frac{1}{\sqrt{L\beta }}\sum \delta b(q)e^{iqx}\text{,} 
\nonumber \\
\frac{1}{\partial _{t}+iB} &=&\frac{1}{\sqrt{L\beta }}\sum g(p)e^{ipx}\text{%
, }g(p)=-\frac{1}{ip_{0}+M\sin p_{1}}
\end{eqnarray}
At zero temperature, the sum over Matsubara frequencies turns into an
integral and one obtains:

\begin{equation}
\delta E_{fluctuations}=\frac{1}{2\beta }\log \det \left( \frac{\delta ^{2}F%
}{\delta B^{2}}\right) \stackrel{T=0}{=}\frac{1}{8\pi ^{2}}\int_{-\infty
}^{\infty }dq_{0}\int_{-\pi }^{\pi }dq_{1}\log F(q_{0},q_{1})
\end{equation}
$F(q_{0},q_{1})$ can be expressed in terms of elementary functions: 
\begin{eqnarray}
F(q_{0},q_{1}) &=&1-\frac{1}{L}\sum_{p}\left( 1-e^{2ip}\right) \frac{%
n_{F}(M\sin (p-\frac{q_{1}}{2}))-n_{F}(M\sin (p+\frac{q_{1}}{2}))}{M\sin (p-%
\frac{q_{1}}{2})-M\sin (p+\frac{q_{1}}{2})-iq_{0}}  \nonumber \\
&=&1+\frac{1}{\pi M\sin (\frac{q_{1}}{2})}\int_{0}^{\frac{q}{2}}dp\frac{\sin
^{2}p\cos p}{\cos ^{2}p+(\frac{q_{0}}{2M\sin (\frac{q_{1}}{2})})^{2}} 
\nonumber \\
&=&1+\frac{1}{\pi M}\left( -1+\frac{\sqrt{1+x^{2}}}{\sin (\frac{q_{1}}{2})}%
\tanh ^{-1}\left( \frac{\sin q_{1}/2}{\sqrt{1+x^{2}}}\right) \right) \text{, 
}x=\frac{q_{0}}{2M\sin \frac{q_{1}}{2}}
\end{eqnarray}
The total energy that includes the fluctuations about the un dimerized
saddle point is, therefore,

\begin{eqnarray}
E_{N} &=&\frac{M^{2}}{8}-\frac{NM}{2\pi }+\delta E_{fluctuations}  \nonumber
\\
\delta E_{fluctuations}(M) &=&\frac{M}{\pi ^{2}}\int_{0}^{\pi
}dq\int_{0}^{\infty }dx\log F\sin \frac{q}{2}dxdq
\end{eqnarray}
Setting $N=3$ and $M=\frac{6}{\pi }$ we find $\delta E_{fluctuations}=0.036$
and since the correction is so small, it is not necessary to do the Gaussian
approximation self consistently. The total energy differs from the exact
Bethe Ansatz result still by much more than 100\%. So the loop expansion
converges very well, but...to the wrong result.

\section{The problem of gauge invariance}

In the Majorana representation of eq(1) , there is an infinity of operators
that commute with the Hamiltonian \cite{CIT} 
\begin{eqnarray}
\text{ }Q(\eta _{x}) &=&i\prod_{\mu =1}^{N}\eta _{x\mu }  \nonumber \\
\lbrack Q(\eta _{x}),H] &=&0  \nonumber \\
N\text{ even} &:&\text{ }\left[ Q(\eta _{x}),Q_{y}(\eta _{y})\right] _{-}=0 
\nonumber \\
N\text{ odd} &:&\text{{}}\left[ Q(\eta _{x}),Q_{y}(\eta _{y})\right]
_{+}=\left( \frac{1}{2}\right) ^{N-1}\delta _{x,y}
\end{eqnarray}
The oscillation in statistics of $Q(\eta _{x})$ as a function of $N$ may be
considered to be responsible for the even/odd oscillation in the energy of a
small cluster as a function of $N$ that was observed in \cite{DFFT}. The $%
\eta _{x}$ are gauge transformed by commuting (anticommuting) them with the
operators $Q(\eta _{x})$. The theory of quantising gauge invariant systems
is highly developed \cite{GaugeReview}, but a caricature of existing methods
runs as follows:

\begin{itemize}
\item  some redundant parameter is eliminated and the system, now without
gauge invariance, is treated in terms of the remaining variables by
conventional methods -

\item  or, alternatively, one refrains from eliminating redundant variables.
Instead, one imposes a constraint on the space of variables and provides a
correction factor in a way that essentially extracts the volume of the gauge
group from the partition function. This is done by inserting unity in the
form of

\begin{equation}
1=\Delta _{FP}(\eta )\int D\varepsilon \delta (f(\eta _{\varepsilon }))
\end{equation}
into the partition function, where $f(\eta )$ is a function of $\eta $ that
breaks the symmetry, and where $D\varepsilon $ integrates over the group 
\cite{GaugeReview}. $\Delta _{FP}(\eta )$ is the determinant of the operator 
$\frac{\delta f}{\delta \varepsilon }$ and is represented by auxiliary or
''ghost '' particles.
\end{itemize}

To apply the first method, one notices that the constant multiplicity $%
2^{L/2}$ of extra spurious states in a system of $L$ spins can be understood
in the representation\cite{Shastry}: 
\begin{equation}
\eta _{x\mu }=\sigma _{x\mu }f_{x}
\end{equation}
where $f_{x}$ are Majorana fermions and $\sigma _{x\mu }$ are Pauli matrices
of the group $O(N)$. It can be shown that the extra fermions $f_{x}$ get
eliminated by imposing a positive spectrum , $i\eta _{x}\eta _{y}>0$, on
a non overlapping covering by dimers $(x,y)$ or on dimers between points on
the original system and an extra copy of it. In the second approach, because
the symmetry group is fermionic, $\Delta _{FP}(\eta )$ is actually the {\em %
inverse} of the determinant of $\frac{\delta f}{\delta \varepsilon }$ and
will be represented by {\em bosonic} auxiliary or ''ghost '' particles. The
details of this gauge approach to the Majorana representation still remain
to be worked out, however.

In conclusion, we have seen that ground state energy of a spin 1/2
antiferromagnetic chain treated by Majorana representation and mean field
plus Gaussian fluctuations comes out incorrectly, although the loop
correction is small. This strongly suggests that a proper treatment of the
Majorana representation of spin systems must address the problem of local $%
Z_{2}$ invariance or the local supersymmetry of the system.

\section{Acknowledgements}

I am indebted to Piers Coleman, Alexei Tsvelik and Diptiman Sen for comments
on their work, Yann Meurdesoif for encouragement and Bernard Bonnier for
reading the manuscript. My most humble thanks, however, go to Jack Donohue
for shouldering my teaching load at the university of Bordeaux.

\end{document}